\documentclass[useAMS,usenatbib,a4paper,11pt,aas_macros]{article}
\pdfoutput=1
\usepackage{graphicx}
\usepackage{url}
\usepackage{slashed}
\usepackage{color}
\usepackage{xspace}
\usepackage[utf8]{inputenc}
\usepackage{amssymb}
\usepackage{natbib}
\usepackage{hyperref}

\newcommand{\be}{\begin{equation}}
\newcommand{\ee}{\end{equation}}
\newcommand{\bea}{\begin{eqnarray}}
\newcommand{\eea}{\end{eqnarray}}
\newcommand{\mdm}{M_{DM}}
\newcommand{\gev}{\mathrm{GeV}}

\newcommand{\sigmavgg}{\sigma v_{\gamma\gamma}}
\newcommand{\sigmavgz}{\sigma v_{\gamma Z}}
\newcommand{\sigmavgh}{\sigma v_{\gamma h}}

\title{\bf Extending Fermi-LAT and H.E.S.S. Limits on  Gamma-ray Lines from  Dark Matter Annihilation
}
\author{Stefano Profumo$^1$\thanks{profumo@ucsc.edu},\, Farinaldo S. Queiroz$^2$\thanks{farinaldo.queiroz@mpi-hd.mpg.de}, \,Carlos E. Yaguna$^2$\thanks{carlos.yaguna@mpi-hd.mpg.de} \\[0.5cm]
{\footnotesize \it $^1$ Department of Physics, University of California, Santa Cruz}\\
{\footnotesize \it Santa Cruz Institute for Particle Physics, 1156 High St, Santa Cruz, CA 95064}
\and
{\small \it $^2$ Max-Planck-Institut f\"ur Kernphysik,}\\
{\small \it Saupfercheckweg 1, 69117 Heidelberg, Germany}
}
\date{}

\begin{document}

\maketitle

\begin{abstract}
\noindent Gamma-ray lines from dark matter annihilation ($\chi\chi\to \gamma X$, where $X=\gamma,h,Z$) are always accompanied, at lower energies, by  a  continuum gamma-ray spectrum stemming both from radiative corrections ($X=\gamma$) and from the decay debris of the second particle possibly present in the final state ($X=h,Z$). This model-independent gamma-ray emission can be exploited to derive novel limits on gamma-ray lines that do not rely on the line-feature. Although  such limits are not expected to be as stringent, they can be used to probe the existence of $\gamma$-ray lines for dark matter masses beyond the largest energies accessible to current telescopes. Here, we use continuous gamma-ray searches from Fermi-LAT observations of Milky Way dwarf spheroidal galaxies and from H.E.S.S. observations of the Galactic Halo to extend the limits on the annihilation cross sections into monochromatic photons to dark matter masses well beyond $500$ GeV (Fermi-LAT) and $20$ TeV (H.E.S.S.). In this large mass regime, our results provide the first constraints on $\gamma$-ray lines from dark matter annihilation.
\end{abstract}

\noindent {\bf Key words:}  astroparticle physics; gamma-rays: general; gamma-rays: galaxies; line: identification; (cosmology:) dark matter


\section{Introduction}
Indirect searches for dark matter are one of the most promising ways to detect the dark matter particle and to determine its fundamental properties \cite{Buckley:2013bha}. Indirect signatures include searches for anomalous $\gamma$, $\nu$, $e^+$ and $\bar p$ emission produced by dark matter annihilation in astrophysical objects that could be observed over, and disentangled from, the expected backgrounds from known astrophysical emission. Gamma-ray searches \cite{Bringmann:2012ez}, in particular, currently provide some of the most stringent and robust constraints on the dark matter annihilation cross section into different final states \cite{Cirelli:2015gux}.

These searches can be generically divided into three different categories: 

(i) Prompt continuum emission, ranging from $E\sim M_{DM}$ down to soft gamma-ray energies, from dark matter annihilation into any SM final state (e.g. $b\bar b$ or $W^+W^-$) radiating photons, and/or producing photons as a result of hadronization of the decay products of the final-state particles (for example $\pi^0\to\gamma\gamma$);

(ii) Secondary continuum emission, resulting from the radiative processes associated with stable, charged particles produced in the annihilation event (such as electrons and positrons); this emission  extends from radio all the way to gamma-ray frequencies (see e.g. \cite{Colafrancesco:2005ji,Colafrancesco:2006he} and Ref.~\cite{Profumo:2010ya} for a review)

(iii) Monochromatic (line-like) photon emission, resulting from two-body final states containing a photon, for example $\gamma\gamma$, $\gamma Z$, and $\gamma h$, or from internal Bremsstrahlung \cite{Bergstrom:2004cy,Bergstrom:2004nr,
Bergstrom:2005ss,Bringmann:2007nk,Bringmann:2011ye}.

While the continuum emission expected from (i) and (ii) can have morphological and spectral features that would enable distinguishing a dark matter component from astrophysical diffuse emission, lines in the GeV-TeV energy range, especially if accompanied by an extended morphology\footnote{Cold pulsar winds can produce gamma-ray lines, but are point-like sources \cite{Aharonian:2012cs}.} \cite{Carlson:2013vka} can be unmistakably associated with dark matter annihilation (or decay). As a result, current limits on the dark matter annihilation cross section into monochromatic photons are often more constraining than those from continuum emissions\footnote{On the other hand, the pair-annihilation cross section of dark matter to two-body final states including a photon is generically suppressed by requiring the dark matter to be essentially electrically neutral.}. 

At a dark matter mass ($\mdm$) of $100$ GeV, for example, the Fermi-LAT constraint on $\sigma v(\chi\chi\to \gamma\gamma)$ --denoted in the following by $\sigmavgg$-- is almost three orders of magnitude stronger than that on $\sigma v(\chi\chi\to b\bar b)$ \cite{Ackermann:2015zua,Ackermann:2015lka}. But while the Fermi-LAT limits on $\sigma v(\chi\chi\to b\bar b)$ extend all the way up to dark matter masses of 10 TeV,  the limits on $\sigmavgg$  are restricted to $\mdm <500~\gev$, which corresponds to the largest photon energy probed by Femi-LAT collaboration in their spectral line \cite{Ackermann:2015lka}. A similar effect occurs in the H.E.S.S. telescope beyond 20 TeV. Hence, we will refer to those energies as the detectors energy limit. 

Consequently, Fermi-LAT data do not provide, at face value, limits  on $\sigmavgg$ for dark matter masses beyond $500$ GeV. Similarly, H.E.S.S. data can only constrain $\sigmavgg$ for dark matter masses smaller than about $20$ TeV \cite{Abramowski:2013ax}. In other words,  the energy upper limit  corresponds to the maximal dark matter mass probed by these telescopes in the search for gamma-ray lines.

Monochromatic photons from dark matter annihilation, however, always produce a continuum gamma-ray spectrum that extends down to low energies. This model-independent spectrum is generated either by radiative emission \cite{Ciafaloni:2010ti} (for $\gamma\gamma$) or by the decay of the particle accompanying the photon (for $\gamma Z$ and $\gamma h$). By exploiting this continuum spectrum, bounds on $\sigmavgg$, $\sigmavgz$ and $\sigmavgh$ may be derived even for dark matter masses much larger than the energy upper limit of a given gamma-ray telescope. This simple observation, which seems to have been overlooked in the previous literature, is the basis of our work. 

Although the  bounds derived from this continuum spectrum are not expected to be as stringent as those based on the line-feature, they allow to extend the limits on gamma-ray lines from dark matter annihilation to masses not previously testable due to the energy range of the telescopes. In this way, available experimental data can be used to constrain in a novel and independent way the fundamental properties of the dark matter particle \cite{Yaguna:2009cy}. 

As an example, consider a 1 TeV dark matter particle pair-annihilating into photon pairs with a $100\%$ branching ratio. At first sight,  Fermi-LAT seems to have  no sensitivity to such a scenario because the signal --a $\gamma$-ray line with $E_\gamma = 1$ TeV-- lies above the Fermi-LAT energy limit. But, the final state photons may radiate, via weak corrections,  $W^{\pm}$ gauge bosons that will in turn give rise to a gamma-ray continuum emission within the Fermi-LAT energy range.  Thus, thanks to the continuum emission associated with the lines, Fermi-LAT becomes  sensitive to such a dark matter particle.

Here, we apply this idea to both Fermi-LAT and H.E.S.S. limits on continuum emission from dark matter.  Specifically, we use Fermi-LAT observations of dwarf spheroidal galaxies using PASS 8 class of events \cite{Ackermann:2015zua}  to set limits on $\sigmavgg$, $\sigmavgz$ and $\sigmavgh$ for dark matter masses above 500 GeV, and H.E.S.S. observations of the Galactic center region \cite{Abramowski:2011hc} to extend the range of limits on $\gamma$-ray lines to dark matter masses beyond  20 TeV. In this mass range, our results provide the first, model-independent constraints on $\gamma$-ray lines from dark matter annihilations. 


\section{Continuum emission associated with gamma-ray lines}

\begin{figure}[!t]
\centering
\begin{tabular}{ccc}
\includegraphics[scale=0.5]{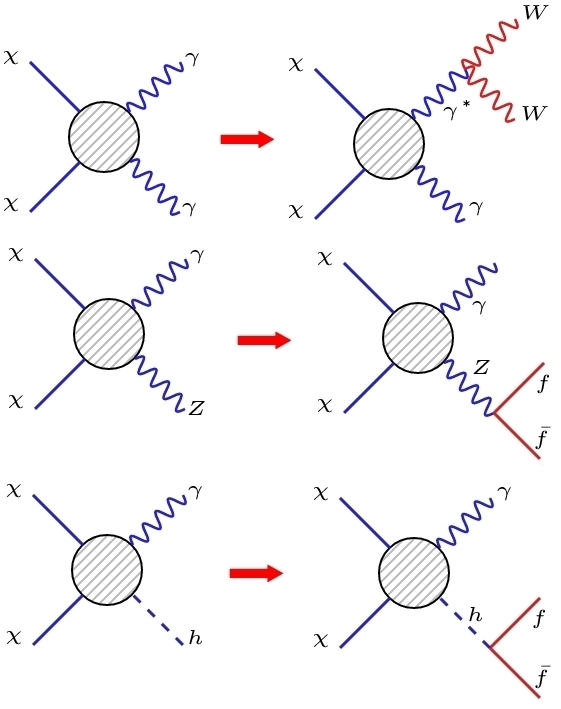}
\end{tabular}
\caption{Diagrams ilustrating how a continuous gamma-ray emission arises from annihilation into gamma-ray lines.\label{fig:diagrams} We acknowledge the use of the online tool available at {\rm http://feynman.aivazis.com/}} 
\end{figure}

Gamma-ray lines from dark matter annihilation are always accompanied by a continuum spectrum at lower energies. For the $\gamma Z$ and $\gamma h$ final states, this spectrum arises mostly from the decay (and subsequent hadronization  of the decay products) of the  $Z$ and $h$ produced in association with the photons. For the $\gamma \gamma$ final state, a continuum emission stems from radiative corrections from QED and weak interactions \cite{Ciafaloni:2010ti}. A final state photon may, for instance, radiate $W$ bosons (see figure \ref{fig:diagrams}), which eventually decay producing a well-known continuum gamma-ray spectrum. For the $\gamma Z$ and $\gamma h$ final states this lower-energy gamma-ray emission is more pronounced since both $Z$ and $h$ decays  lead to sizable continuum gamma-ray emissions, as exemplified in Fig.\ref{fig:diagrams}.

It is important to note that the continuum spectrum associated with $\gamma$-ray lines is model-independent, for it arises entirely from Standard Model physics.  Therefore, such a spectrum can be used to probe the existence of the primary annihilation channel containing the photon. Here, we use the  gamma-ray fluxes as incorporated into the PPPC Code \cite{Cirelli:2010xx}, which already include the relevant electroweak corrections \cite{Ciafaloni:2010ti}.

\begin{figure}[tb]
\includegraphics[scale=0.42]{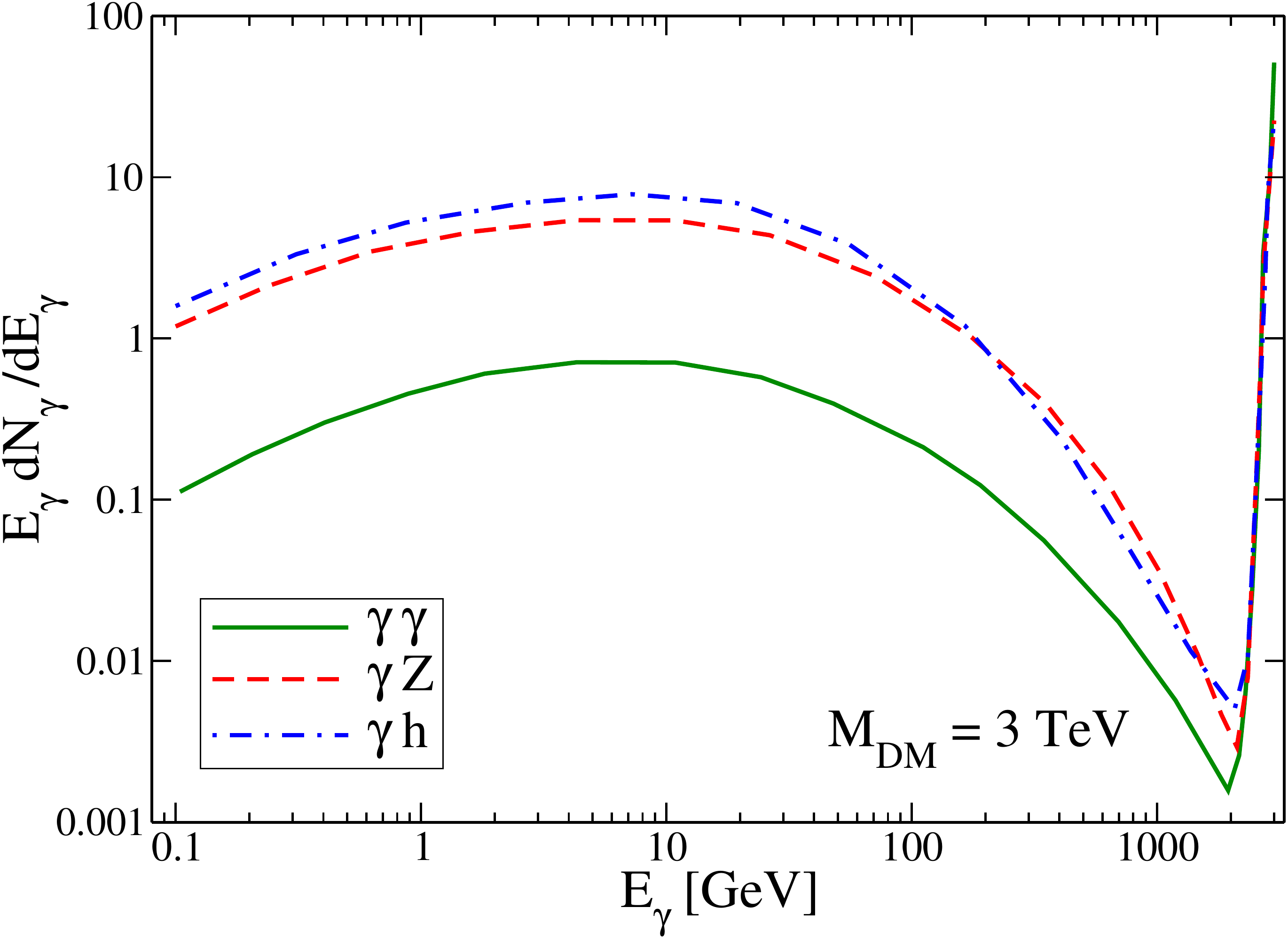}
\caption{The continuum spectrum associated with gamma-ray lines for a dark matter mass of 3 TeV. The primary annihilation channel is assumed to be $\gamma\gamma$ (solid blue line), $\gamma Z$ (dashed orange line), or $\gamma h$ (dash-dotted green line).   \label{fig:dnde}} 
\end{figure} 

For the sake of illustration, we show, in  figure \ref{fig:dnde}, the continuum spectrum from the annihilation of dark matter into   final states containing one or two monochromatic photons. The dark matter mass is assumed in the figure to be  3 TeV, whereas the primary annihilation channels were taken to be $\gamma\gamma$ (solid green line), $\gamma Z$ (dashed red line), or $\gamma h$ (dash-dotted blue line, with $h$ assumed to be a purely Standard Model-like Higgs). Notice that the shape of the spectrum is similar for all three channels but, as expected, the $\gamma Z$ and $\gamma h$ final states produce many more photons than the $\gamma\gamma$ one, which is generically suppressed by factors of order $\sim\alpha_2\ln^2 \mdm^2/M_W^2$ \cite{Ciafaloni:2010ti}. For this particular dark matter mass, the difference between the $\gamma\gamma$ continuum emission and that from $\gamma Z$ and $\gamma h$ amounts to  one order of magnitude at most. The $\gamma Z$ and $\gamma h$ final states give rise to relatively similar spectra due to the similar gamma-ray yield resulting from the $Z$ and $h$ decays.

For dark matter annihilating  into monochromatic photons, the differential flux of photons from a given angular direction $\Delta\Omega$ is given by
\begin{equation}
\frac{d\Phi_\gamma(\Delta \Omega)}{dE}(E_\gamma)=\frac{1}{4\pi}\frac{\sigma v_{\gamma X}}{2 M_{DM}^2}\frac{dN_\gamma}{dE_{\gamma}}\cdot J_{ann}
\end{equation}where $J_{ann}$ is the annihilation J-factor, 
\begin{equation}
   J_{ann} \; =\; \int_{\Delta \Omega} d\Omega \int \rho_{DM} (s)\, ds
   \quad ,\quad s\; =\; s(\theta )\quad ,
 \label{eq:Jfac_def}
\end{equation}where $\rho_{DM}$ is the dark matter density which is assumed be well described by a NFW profile,

\begin{equation}
   \rho_{DM} (r) \; =\; \frac{\rho_s}{r/r_s (1+ r/r_s)}\quad .
 \label{eq:DM_profile}
\end{equation}where $\rho_s$ and $r_s$ are the characteristic density and scale
radius determined dynamically from the maximum circular velocity $v_c$ and the enclosed mass contained up to the radius of maximum $v_c$  as discussed in \cite{Ackermann:2013yva}. The integral is performed over the line of sigh element 
within the solid angle $\Delta\Omega$. In this work we adopted the J-factors listed in the table I of \cite{Ackermann:2015zua} which is equivalent to \cite{Ackermann:2013yva}. As for $\frac{dN_\gamma}{dE_{\gamma}}$ is the differential $\gamma$-ray yield per annihilation into the  final state $\gamma X$, with $X=\gamma, Z, h$. Since this yield is model-independent --see figure \ref{fig:dnde}--  gamma-ray data can be used to constrain regions in the plane $M_{DM}$ vs $\sigma v_{\gamma X}$.

In our analysis we derive limits from Fermi-LAT and H.E.S.S. data.

\section{Fermi-LAT limits}


The Fermi-LAT telescope has observed a wealth of dwarf spheroidal satellite galaxies of the Milky Way, which are known to be dark matter-dominated objects. The resulting data belonging to the P8R2SOURCEV6 event class, based on  six years of observations, span energies between 500 MeV and 500 GeV. The collaboration makes use of the Pass-8 software which results into an improved point spread function and effective area compared to previous releases and includes the third point source Fermi-LAT catalog (3FGL). Since none of the dwarf galaxies presents any significant gamma-ray excess, tight constraints on the dark matter annihilation cross section can be placed \cite{Ackermann:2015zua} --see also \cite{Ahnen:2016qkx,Ahnen:2016qkx,Oman:2016zjn,Li:2015kag,Dutta:2015ysa,Zitzer:2015eqa,
Rico:2015nya,Bonnivard:2015tta,Bonnivard:2015xpq,Hooper:2015ula,Geringer-Sameth:2015lua,Drlica-Wagner:2015xua,Baring:2015sza,Queiroz:2016zwd}.

\begin{figure}[!t]
\includegraphics[scale=0.4]{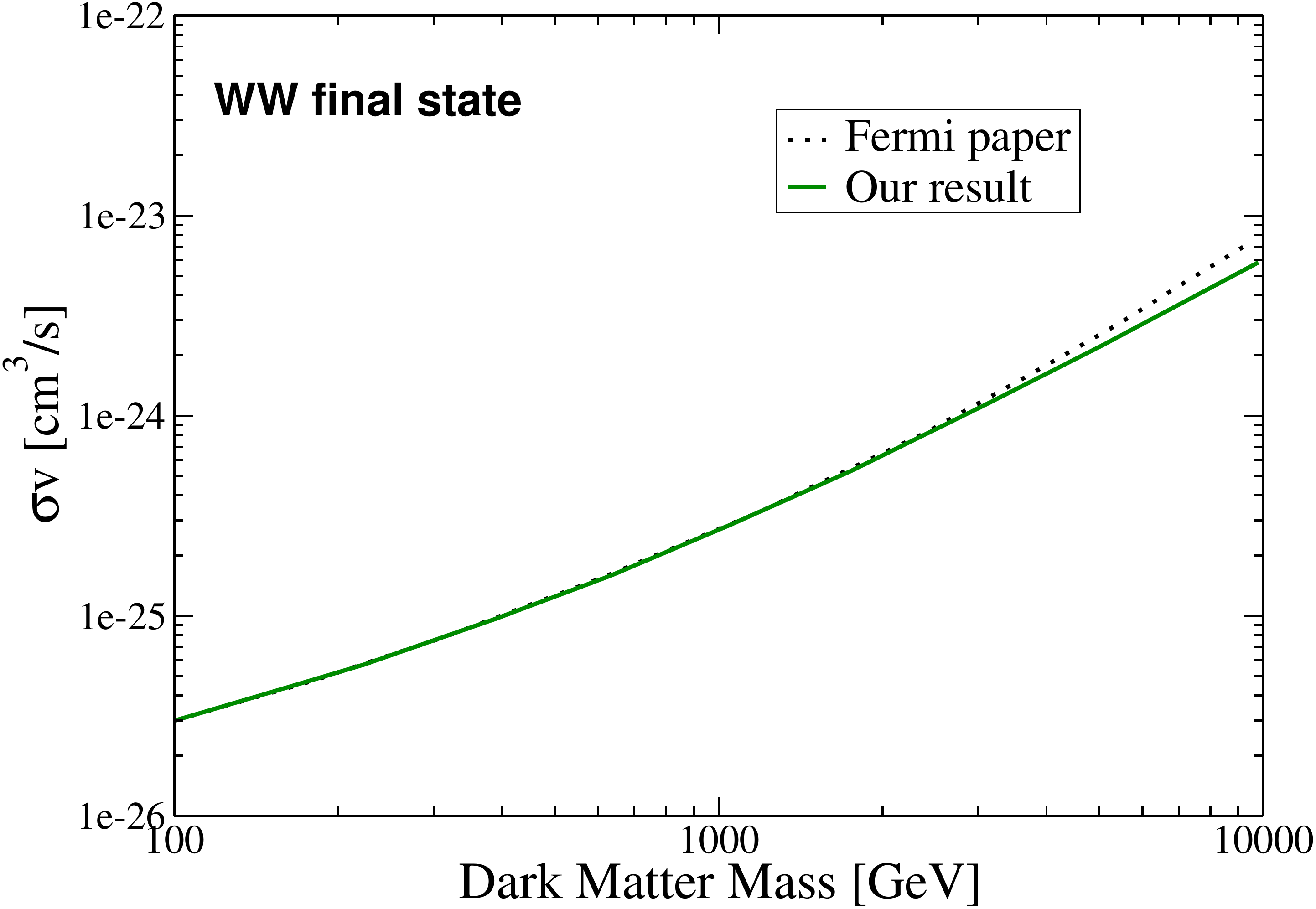}
\caption{\small Comparison between Fermi-LAT official limit and ours for annihilation into WW. It is clear that our limits have a very good agreement wth the official limits from Fermi-LAT. Similar conclusions are found for all channels.\label{Fermipapercompplot}} 
\end{figure}

Fermi-LAT has made publicly available\footnote{http://www-glast.stanford.edu/pub\_data/1048/} binned Poisson maximum-likelihood tools that can be used to reproduce the individual limits stemming from any of the dwarf galaxies in the study, without accounting for uncertainties on the J-factors of each dwarf galaxy, assumed to be described by a NFW dark matter density profile. Here we perform a joint likelihood analysis across the 15 dwarf galaxies, and treat the J-factor as a nuisance parameter, following the recipe described in the supplemental material of Ref.~\cite{Ackermann:2015zua}. In doing so, we use the likelihood of an individual dwarf i, defined as,
\begin{equation}
   \tilde{\mathcal{L}_i}(\mu,\theta_i = 
   \lbrace\alpha_i,J_i\rbrace |D_i)=\mathcal{L}_i(\mu,\theta_i|D_i)\mathcal{L}_J (J_i|J_{obs,i},\sigma_i)
\end{equation}where $\mu$ encompasses the parameters of the DM model, i.e. the ratio of the dark matter annihilation cross section and mass, whereas $\theta_i$ refers for the set of nuisance parameters from the LAT analysis ($\alpha_i$) and J-factors of the dwarf galaxies $J_i$, with $D_i$ being the gamma-ray data. The former is provide by the Fermi-LAT team as mentioned. In order to account for the statistical uncertainties on the J-factors of each dwarf galaxy, a J-factor likelihood function is defined as follows,

\begin{eqnarray}
\mathcal{L}_J (J_i| J_{obs,i}, \sigma_i) &= & \frac{1}{ln(10)J_{obs,i}\sqrt{2\pi} \sigma_i}\nonumber\\
& \times  & \exp \left\{ -\frac{(log_{10}(J_i)-log_{10}(j_{obs,i}))^2}{2\sigma_i^2} \right\}\nonumber
\end{eqnarray}where $J_i$ is referred as the true value of the J-factor of a dwarf galaxy $i$, whereas $J_{obs,i}$ is the measured J-factor with error labelled as $\sigma_i$. Then we join the likelihood terms,
\begin{eqnarray}
\mathcal{L}_i (\mu, \theta_{i}| D_i) = \prod_j \mathcal{L}_i (\mu, \theta_i | D_{i,j})
\label{Maxlike}
\end{eqnarray}and perform a test statistic (TS), with $TS= -2 ln ( \mathcal{L} (\mu_0,\widehat{\theta}|D)/ \mathcal{L} (\widehat{\mu},\widehat{\theta}|D))$, which gives rise to 95\% C.L. upper limit on the energy flux as explained in \cite{Rolke:2004mj} by imposing a change in the log-likelihood of $=2.71/2$ from its maximum. In order to demonstrate that our results are solid, we reproduced Fermi-LAT limits from the usual channels finding a very good agreement. See figure \ref{Fermipapercompplot} where we show that explicitly for the $W^+W^-$ channel.

With the maximum likelihood at hand, we feed a given gamma-ray spectrum, here the low-energy continuum from $\gamma\gamma$,$\gamma Z$ and $\gamma h$, and we obtain a bound on the annihilation cross section arising from the gamma-ray continuum emission from a stack of 15 dwarfs. 

\begin{figure}[!t]
\includegraphics[scale=0.43]{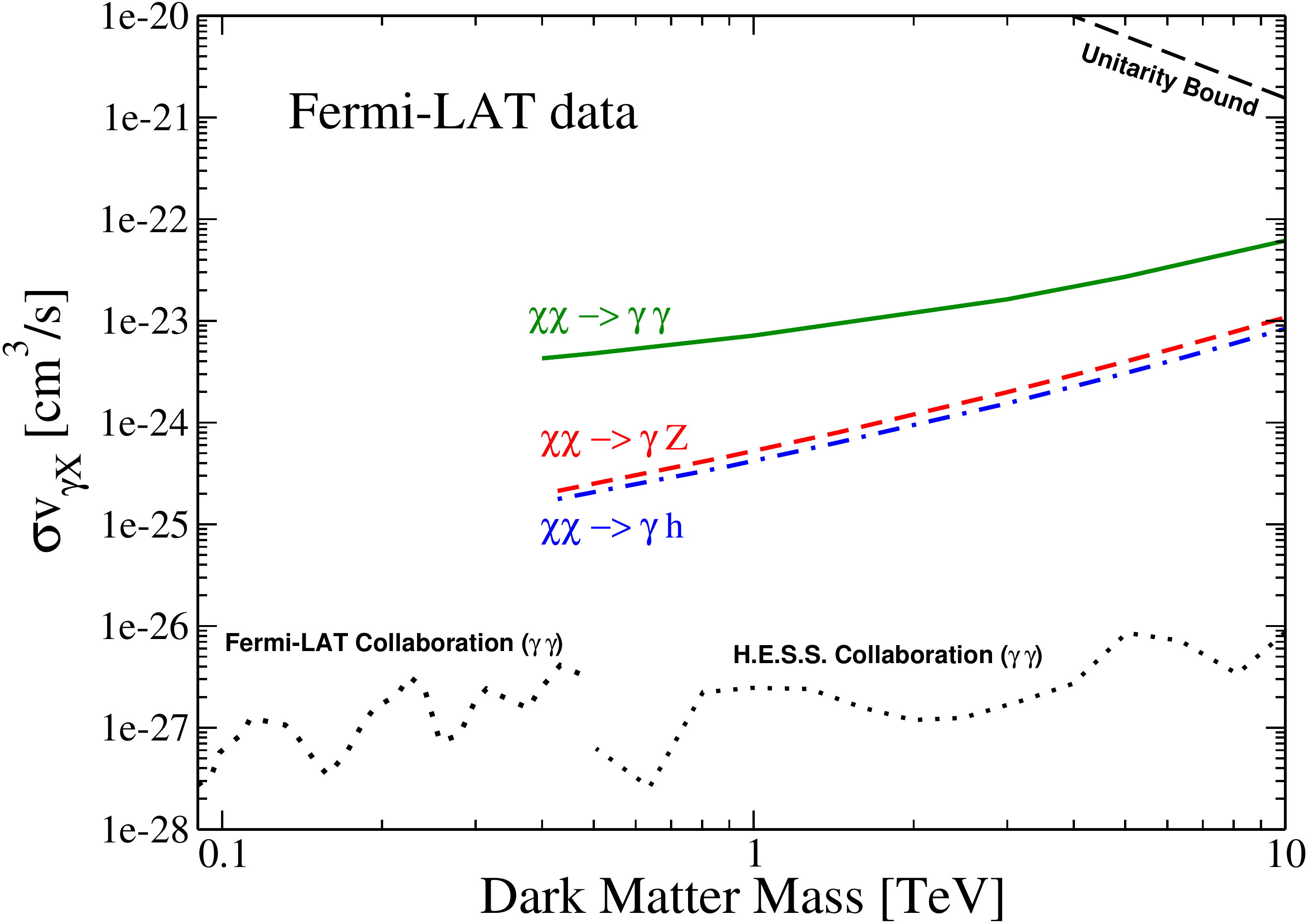}
\caption{\small Limits on $\gamma$-ray lines from dark matter annihilation derived from Fermi-LAT data.  The colored lines show our results --based on the continuum spectrum and using observations of dwarf galaxies-- for different cross sections: $\sigmavgg$ (solid green line), $\sigmavgz$ (dashed red line), and $\sigmavgh$ (dash-dotted line). The dotted black line shows instead the limits derived by the Fermi-LAT collaboration \cite{Ackermann:2013uma}--based on the line-feature and using data from the region around the Galactic center. For reference, the unitarity bound on s-wave annihilation cross sections is also shown (dashed black line; see Ref.~\cite{Griest:1989wd} for details on the unitarity bound and the associated assumptions).  \label{fig:Fermi}} 
\end{figure}

Our results are summarized in Fig.~\ref{fig:Fermi}. It shows the $95\%$ C.L. limits on the thermally averaged, zero-temperature pair-annihilation cross section to $\gamma X$, for $X=\gamma,Z,h$. As previously discussed, the annihilation final states $\gamma Z$ and $\gamma h$ yield similar gamma-ray emission, and consequently  nearly equivalent limits, whereas pure annihilation into photon pair is less constraining due to the lower associated continuum emission level. For reference, we display in this figure the current bounds on gamma-ray lines reported by the Fermi-LAT \cite{Ackermann:2015lka} and H.E.S.S. collaborations (dotted black lines).

In addition,  the unitarity bound on the size of the pair-annihilation cross section \cite{Griest:1989wd} (dashed black line) is also shown. This bound, derived via partial-wave unitarity, sets a model-independent upper limit on the annihilation rate of the dark matter particles.  Today, this upper limit amounts, for s-wave annihilation, to $\sigma v \sim \frac{1.5 \times 10^{-13} cm^3 s^{-1}}{ (GeV/m_{DM})^2}$ \cite{Beacom:2006tt}, where $v\approx10^{-3}$ was used for the velocity of the dark matter particles. Only the region below the dashed black line is thus consistent with unitarity. 


Our new limits, though not as stringent as those based on the line-feature, provide the first Fermi-LAT constraints on gamma-ray lines from dark matter annihilation for $\mdm>500$ GeV.  Even if these limits are weaker than (or consistent with) the current  H.E.S.S. limits \cite{Abramowski:2013ax},  they constitute a new and independent limit on gamma-ray lines that is based  on a different astrophysical target  and that relies entirely on  Fermi-LAT data.

\section{H.E.S.S. limits}


\begin{figure}[!t]
\includegraphics[scale=0.4]{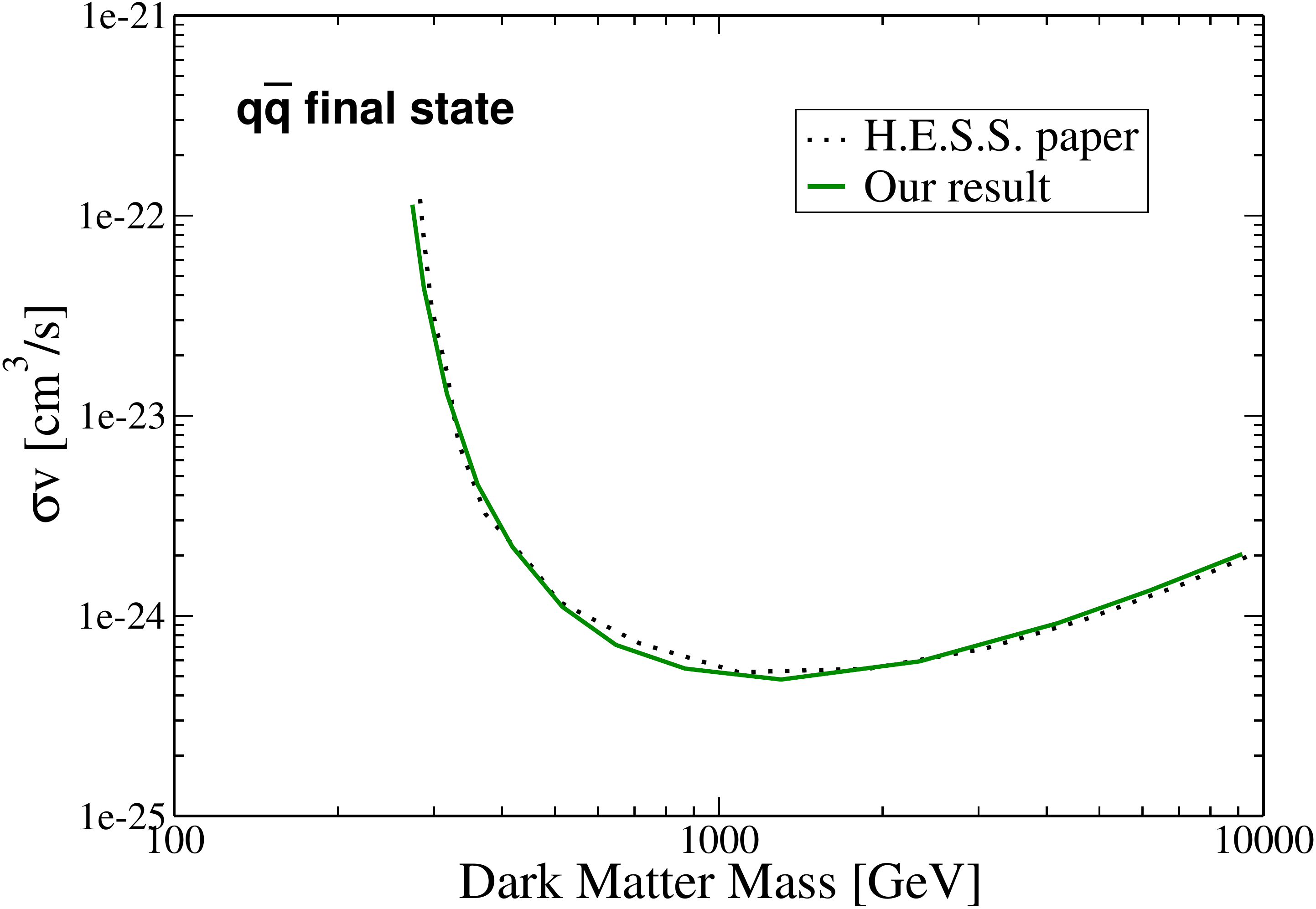}
\caption{\small Comparison between H.E.S.S. official limit and ours for annihilation into quarks using the parametric spectrum defined in \cite{Hill:1986mn,Tasitsiomi:2002vh}. This parametric energy spectrum includes dark matter annihilation into all quark flavors. Our results strongly overlap proving the robustness of our analysis.\label{Hesspapercompplot}} 
\end{figure} 

In \cite{Abramowski:2011hc}, the H.E.S.S. collaboration presented their latest results on the search for a dark matter annihilation signal from the Galactic Center Halo, which are based on  112 hours of GC observations taken over a period of about four years. In their analysis, H.E.S.S. divides the sky into background and source regions, with the former located further away from the Galactic plane, where a dark matter signal is expected to be dimmer. The source region used in \cite{Abramowski:2011hc} was a circular region of radius $1^\circ$ from which Galactic latitudes $|b|<0.3^\circ$ were removed to reduce the possible contamination of the dark matter signal by local $\gamma$-ray sources.  Since no excess emission was found by H.E.S.S., restrictive limits were placed on the dark matter annihilation cross section, which are among the most stringent for dark matter masses above $\sim 800$~GeV. 

\begin{figure}[!t]
\includegraphics[scale=0.43]{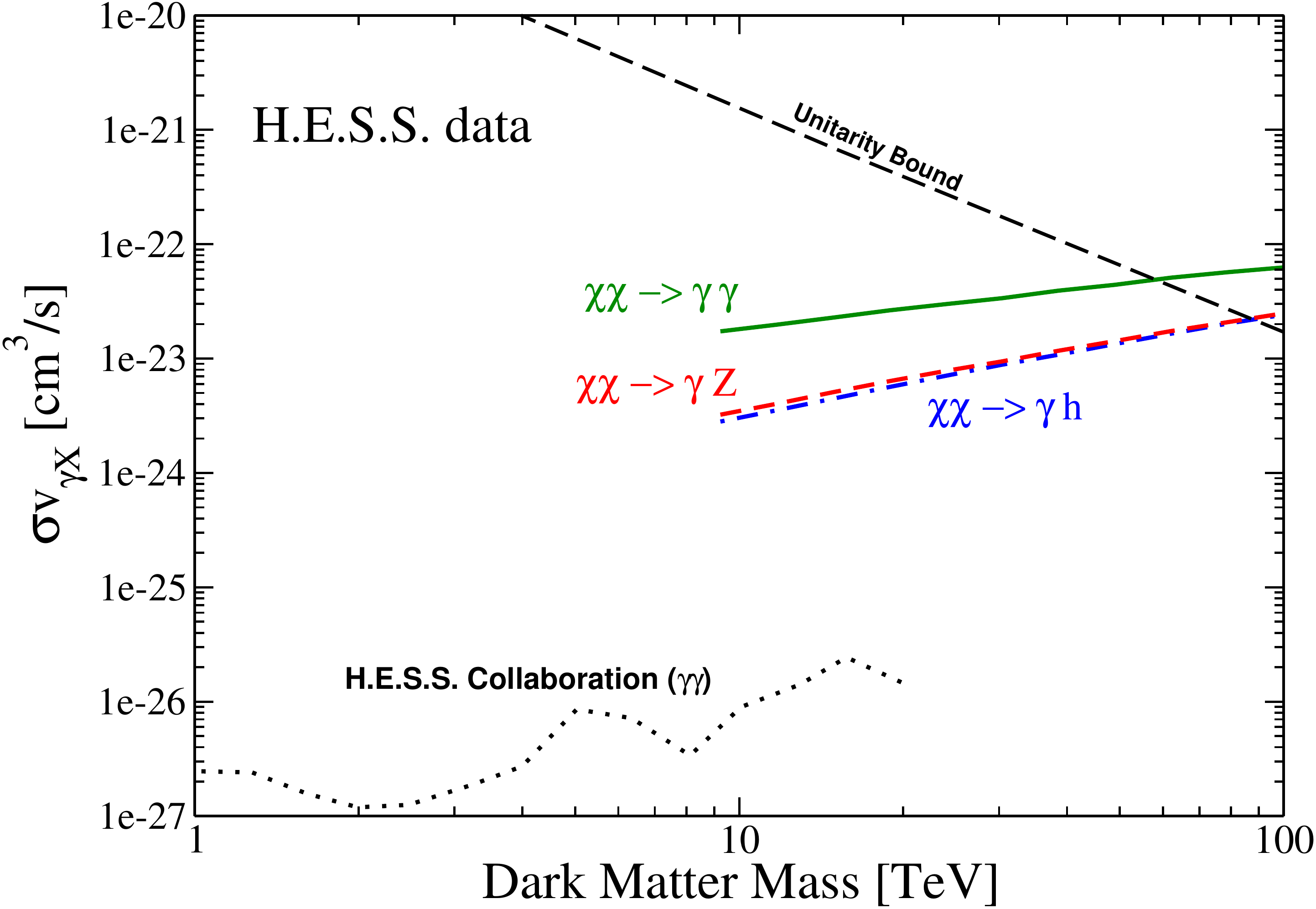}
\caption{\small Limits on $\gamma$-ray lines from dark matter annihilation derived from H.E.S.S. data.  The colored lines show our results --based on the continuum spectrum and using observations of the Galactic center halo-- for different cross sections: $\sigmavgg$ (solid green line), $\sigmavgz$ (dashed red line), and $\sigmavgh$ (dash-dotted line). The dotted black line shows instead the limits derived by the H.E.S.S. collaboration --based on the line-feature and using data from the region around the Galactic center \cite{Abramowski:2013ax}.  For reference, the unitarity bound on the annihilation cross section is also shown (dashed black line). \label{fig:Hess}} 
\end{figure} 

By integrating Fig.3 of Ref.~\cite{Abramowski:2011hc} multiplied by observation time, effective area, and J-factor we can compute the number of signal events for the ON  ($N^{S_{ON}}$) and OFF ($N^{S_{OFF}}$) regions as well as the number of background ($N^B$) events for the OFF regions. ON region refers to the $1^\circ$ circular region around the Galactic Center, whereas the OFF regions are the represented by annulus of radii of $1^\circ$ and $1.5^\circ$  as delimited in Fig.2 of \cite{Abramowski:2011hc} \footnote{The J-factors are given in the H.E.S.S. paper and the effective area was obtained using the code gammapy code \url{https://gammapy.readthedocs.io/en/latest/}, which offers quite good agreement with the effective area made public in several PhD thesis from H.E.S.S. members (e.g. \url{https://www.physik.hu-berlin.de/de/eephys/HESS/theses/pdfs/ArneThesis.pdf})}. We can then define likelihood functions as described in \cite{Lefranc:2015vza} and compute 95\% C.L. limits similarly to the procedure discussed for Fermi-LAT.  In Fig.\ref{Hesspapercompplot} for concreteness we compare our limit with the one obtained by H.E.S.S. collaboration. The limits are for dark matter annihilations into all quarks using the parametric energy spectrum as defined in \cite{Hill:1986mn,Tasitsiomi:2002vh}.  From Fig.\ref{Hesspapercompplot} it is clear that our analysis finds a very good agreement with H.E.S.S. result.

After this cross-check, we computed the 95\% C.L. bound on annihilation cross section vs dark matter mass plane for the channels $\gamma Z$,$\gamma h$ and $\gamma \gamma$, using a NFW profile, similarly to \cite{Abramowski:2011hc}. Our results are summarized in Fig.~\ref{fig:Hess}. Since a 100 TeV dark matter particle annihilating into $\gamma \gamma$ still produces some appreciable amount of continuum gamma-rays below 20 TeV, H.E.S.S. can still probe the properties of such a dark matter particle. To the best of our knowledge, these are the first limits on gamma-ray lines from dark matter annihilation for $\mdm>20$ TeV. 

While the new limits we have derived, both from Fermi-LAT and H.E.S.S., on the dark matter annihilation cross sections into monochromatic photons are above the typical cross sections needed to produce dark matter in the early universe as a thermal relic, they are applicable for example to dark matter models where the production mechanism is non-thermal \cite{Gelmini:2006pq}, or where a modified expansion rate occurs at the time of decoupling (see e.g. \cite{Profumo:2003hq} and references therein). These two possibilities are especially interesting in the large-mass region that our limits cover.

\section{Conclusions}
In this study we demonstrated that the continuum spectrum associated with $\gamma$-ray lines from dark matter annihilation  can be used, in conjunction with currently available data,  to set new constraints on the dark annihilation cross section into monochromatic photons over a very broad range of dark matter particle masses. This continuum spectrum is model independent, and arises from radiative corrections or from the decay of the particle accompanying the photon --a Higgs boson or a $Z$. Using limits from six years of Fermi-LAT observations of local dwarf galaxies with PASS-8, we extended current Fermi-LAT limits on $\sigmavgg$, $\sigmavgz$ and $\sigmavgh$ to dark matter masses well beyond $500$ GeV (Fermi-LAT energy upper limit); similarly, using  H.E.S.S. observations of the Galactic Halo, we extended the corresponding  H.E.S.S. limits to dark matter masses larger than  $20$ TeV. At these very large masses, our limits provide  the first constraints on gamma-ray lines from dark matter annihilation.

\section*{Acknowledgements}
We thank Fermi-LAT Collaboration for the public data and tools used in this work. We are also grateful to the PPPC team for making their code publicly available. We thank Christoph Weniger for discussions. CY is supported by the Max Planck Society in the project MANITOP. SP is partly supported by the US Department of Energy, Contract DE-SC0010107-001.

\bibliographystyle{mnras}
\bibliography{darkmatter}

\end{document}